\documentclass[reprint,prl,twocolumn,superscriptaddress]{revtex4-1}
\usepackage[english]{babel}
\usepackage{ucs}
\usepackage[utf8x]{inputenc}
\usepackage{amsmath}
\usepackage{amsfonts}
\usepackage{amssymb}
\usepackage{graphicx}
\usepackage{bbm}
\usepackage{empheq}
\usepackage{color}

\begin{document}
	
	\title{Magnetic gradient free two axis control of a valley spin qubit in SiGe}
	
	\author{Y.-Y. Liu}
	\author{L. A. Orona}	
	
	\affiliation{Department of Physics, Harvard University, Cambridge, Massachusetts  02138, USA}	
	\author{Samuel F. Neyens} 
	\author{E. R. MacQuarrie} 
	\author{M. A. Eriksson}
	\affiliation{University of Wisconsin-Madison, Madison, 53706, Wisconsin, USA}
	
	\author{A. Yacoby}
	\affiliation{Department of Physics, Harvard University, Cambridge, Massachusetts  02138, USA}
	\date{\today}

	\begin{abstract}
	Spins in SiGe quantum dots are promising candidates for quantum bits but are also challenging due to the valley degeneracy which could potentially cause spin decoherence and weak spin-orbital coupling. In this work we demonstrate that valley states can serve as an asset that enables two-axis control of a singlet-triplet qubit formed in a double quantum dot without the application of a magnetic field gradient. We measure the valley spectrum in each dot using magnetic field spectroscopy of Zeeman split triplet states. The interdot transition between ground states requires an electron to flip between valleys, which in turn provides a g-factor difference $\Delta g$ between two dots. This $\Delta g$ serves as an effective magnetic field gradient and allows for qubit rotations with a rate that increases linearly with an external magnetic field. We measured several interdot transitions and found that this valley introduced $\Delta g$ is universal and electrically tunable. This could potentially simplify scaling up quantum information processing in the SiGe platform by removing the requirement for magnetic field gradients which are difficult to engineer.
		
	\end{abstract}
	
	\maketitle
	
	\section{Intro} 
	\label{sec:intro}
	 Quantum algorithms have been proposed to solve problems that are formidable for classical computers \cite{somma2008quantum, carleo2017solving, aspuru2005simulated, shor1999polynomial} but require quantum hardware to be implemented. Electron spins confined to gate defined semiconductor quantum dots hold promise as quantum bits (qubits) due to the promise of long coherence times and the localized nature of their control, making them promising for scaling up to the large number of qubits required for real algorithms \cite{Golovach2006, Morton11, vandersypen2017interfacing, Maune2012}. These qubits have demonstrated fast initialization, high fidelity readout\cite{Petta2004, Simmons2011}, and the possibility of operating at temperatures above 1 Kelvin \cite{Petit2020, Yang2020}. Single spin qubits have demonstrated high fidelity single qubit gates of 99.9\% \cite{Takeda2018} and two qubit gates above 98\%  \cite{Huang2019}. While the difficulty in addressing single spin quibts might be an obstacle for scaling up, singlet-triplet qubits have the advantage of having a Hamiltonian whose magnitude and direction can be electrically tuned from the exchange energy axis to an additional control axis generated by a {magnetic field gradient}, $\Delta B_Z$  \cite{Foletti2009, Wu2014}. 
	 
	Among all semiconductor platforms, silicon/silicon-germanium (Si/SiGe) is appealing because its weak nuclear spin background minimizes the decoherence caused by magnetic field fluctuations \cite{hanson2007spins, Maune2012}. Moreover, weak spin-orbit coupling further reduces the spin relaxation caused by charge fluctuation. This, however, is a double edged sword, because spin-orbital coupling can also be used for electrical spin control. In the absence of full electrical control, other works have used micromagnets placed near the qubit to create a local magnetic field gradient for an effective spin drive  \cite{Wu2014, vandersypen2017interfacing, Zajac2018, Takeda2018}. Strong field gradients, however, are hard to create over large areas of the sample, posing challenges for scaling up this scheme. Another challenge to developing a high fidelity spin qubit in Si/SiGe is the valley degeneracy, which may also contribute to spin decoherence \cite{Yang2013}. This degeneracy can be measured using magneto spectroscopy, Hall Bar measurements and microwave readout in hybrid cavity quantum electrodynamic (cQED) systems \cite{Borselli2011, Goswami2007, Mi2017} and can be removed by lattice strain and electrostatic confinement \cite{Goswami2007, Friesen2010}. 
	
	In this work we investigate a valley-assisted spin qubit formed in SiGe double quantum dots which does not require a magnetic field gradient to achieve full two axis single qubit control. For interdot transitions where the electron number is (4n,4m) - (4n$\pm$1,4m$\mp$1), where n and m are integers, valley flipping is required for transitions between ground states in each dot. We use a spin funnel scheme  \cite{Petta2004} to map the valley spectrum in each dot. The valley splitting can be tuned using gate controlled electric fields \cite{Goswami2007, Yang2013, Hollmann2020}.  Here we show that valley flipping between the interdot transition provides a g-factor difference and generates a $\Delta B_Z$ rotation whose precession  rate increases linearly with increasing magnetic field. Importantly, this valley introduced g-factor gradient is also electric field dependent. The combined dependence on the magnetic and electrical fields enables a tunable $\Delta B_Z$ rotation. This could potentially simplify scaling up quantum information processing as it removes the need for an external magnetic field gradient.

	\section{Device and method}
	\label{sec:rf_reflectometry}

	\begin{figure*}
		\includegraphics[width = 2\columnwidth]{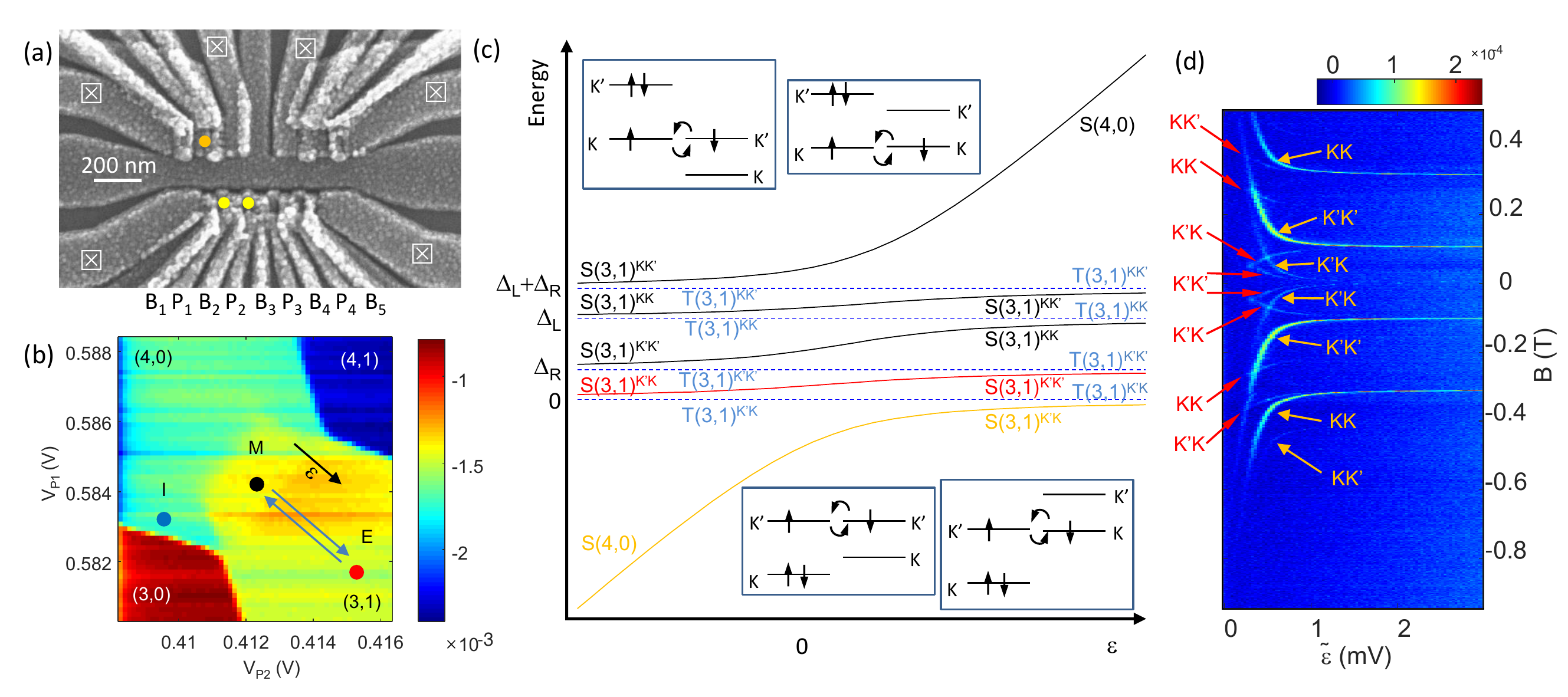}
		\caption{(a) SEM of an overlap style device. The DQD and sensor location are labeled as yellow and orange dots, respectively. (b) Charge stability diagram measured by performing charge readout as a function of plunger gate voltages after being prepared in a random (3,1) state. Spin blockade will keep triplet in configuration (3,1)  and singlet in configuration (4,0) in the bright area in the (4,0) region, giving rise to the readout spot $M$. The spot $I$ is for the initialization of a singlet and spot $E$ is for qubit gate operation.  (c) Schematic level diagram as a function of $\epsilon$ with the singlet plotted as solid lines and triplets plotted as blue dashed lines. Inset: four possible (3,1) states interact with $S_{40}$. (d) Spin funnel obtained by measuring the cross over of singlet to triplet states as a function of $B$ and $\tilde{\epsilon}$. Two sets of curvatures are observed, as emphasized by the red and yellow arrows. The valley states of the triplets are labeled accordingly.}
		\label{fig: device}
	\end{figure*}
	
	Figure\ \ref{fig: device}(a) shows a scanning electron microscope image of a typical device that utilizes an overlap gate geometry to achieve quantum dot confinement \cite{Zajac2015}. The barrier gates, B, create potential barriers for controlling tunneling rates and plunger gates, P, select the charge state in each dot and tune their chemical potential. For this experiment, a double quantum dot (DQD, marked in yellow) is formed at the left two plunger gates $P_1$ to $P_2$, while $P_3$ to $B_5$ control the tunneling rate to a fermi sea. We also form a sensor dot (marked in orange) to perform charge detection and use RF-reflectometry for fast readout.
	
	Figure. \ref{fig: device}(b) demonstrates Pauli-blockade at the (4,0)-(3,1) transition, as required for forming a singlet-triplet qubit. 
	The energy states of the DQD are dependent on which of the two valley eigenstates, K or K', the electrons occupy. {Here we note that valley states K and K’ can be different between dots, and thus there is no orthogonality, which allows valley states to flip during interdot transitions.}  In the (4,0) charge state the four spin valley combinations of the ground orbital state are completely filled. The spin blockade region shown in Fig.\ \ref{fig: device}(b) is cutoff at low $\epsilon$ by transitions into the excited orbital state, which is 200 $\mu$eV higher than the ground state and allows the triplet states to decay into (4,0) states.  This energy is large enough to be ignored in the spin dynamics discussed below.  
	
	The insets of Fig.\ \ref{fig: device}(c) show the four possible (3,1) states that the ground (4,0) state can transition to without a spin flip. We use the notation $\rm (3,1)^{\rm ij}$ where the superscript i represents the valley of the vacancy in the left dot and j represents the valley of the electron in the right dot. The ground (3,1) charge state is then $\rm (3,1)^{\rm K'K}$ and the transition from (4,0) to $\rm (3,1)^{\rm K'K}$ requires flipping an electron from the K' valley to the K valley, suggesting that there is a valley difference between these two states. Assuming that the valley splitting in the left(right) dot is $\Delta_{L(R)}$, the three excited (3,1) valley states would be $\rm (3,1)^{\rm K'K'}$, $\rm (3,1)^{\rm KK}$ and $\rm (3,1)^{\rm KK'}$ and are $\Delta_L$, $\Delta_R$ and $\Delta_L + \Delta_R$ higher in energy compared to the $\rm (3,1)^{\rm K'K}$ state. 	
	
	Figure\ \ref{fig: device}(c) plots the energy diagram of all relevant states as a function of detuning $\epsilon$ between the chemical potential of ground (4,0) and (3,1) charge states. 
	The interdot coupling $t_c$ opens up the avoided crossing of all (4,0)-(3,1) singlet transitions, and leads to the hybridization of different  S$\rm (3,1)^{\rm ij}$ singlets when $2t_c \gtrsim \Delta_{L(R)}$ as shown by the solid lines in Fig.\ \ref{fig: device}(c). For all (3,1) triplet states, the Pauli blockade forbids the interdot transition and give rise to the energy levels as shown by the blue dashed lines. An external magnetic field $B_z$ will further introduce Zeeman splitting $E_Z = \pm g\mu_BB_z$ that lift the degeneracy of triplet states (not shown in the figure).

	\section{Spin Funnel} 
	\label{sec: Valley states spectrum}
	
	Spin funnel measurements can extract the exchange energy $J(\epsilon)$ between the singlet and zero angular momentum triplet state, $T_0$, by detecting the S-$T_+$ degeneracy as a function of magnetic field $B$ and $\epsilon$ \cite{Petta2004}. This allows us to map $J(\epsilon)$ to the Zeeman energy $E^c_Z =g\mu_BB^c_z$. In order to detect this degeneracy, we first prepare the DQD in a singlet by loading the ground (4,0) state at the spot $I$ in Fig.\ \ref{fig: device}(b) and then park at spot M. We then  abruptly pulse the DQD to the exchange location $E$ and evolve for a time of $\tau_E =$ 1 $\mu$s.  Finally, we perform readout of the spin state at position M. During the process we keep the pulse along the diagonal ($\epsilon$) direction such that $\Delta V_{P2} = -\Delta V_{P1} = \tilde{\epsilon}$ with a lever arm that converts from gate potential to energy given by $\epsilon = 0.25 \tilde{\epsilon}$ meV/mV. At a detuning where $E_Z = J(\epsilon)$, an interaction between the singlet state and the degenerate $T_+$ state results in a finite probability $P_T$ of flipping the state into a triplet state \cite{Foletti2009, Petta2008}. Fig.\ \ref{fig: device}(d) demonstrates the funnel data over $\tilde{\epsilon} = $ 0 - 3 mV and a magnetic field range $B =$ -0.9 $\sim$ 0.5 T. A multitude of spin funnel features symmetric in magnetic field are clearly observed indicating multiple triplet branches become degenerate with {different singlet states}. 
	
	The orange arrows point to 4 pairs of spin funnels where $E^c_z(\tilde{\epsilon})$ is quickly changing with $\tilde{\epsilon}$. The 1st pair approaches zero field at large $\tilde{\epsilon}$ and maps out the the ground state singlet energy that is plotted as  the orange line in Fig.\ \ref{fig: device}(c). This results from the crossing of this ground singlet state and $T_+(3,1)^{\rm K'K}$. The other 3 pairs are parallel to the first spin funnel and offset by 12 $\mu$eV, 33 $\mu$eV and 45 $\mu$eV, which corresponds to the intersection between the ground singlet and $T_+(3,1)^{\rm K'K'}$, $T_+(3,1)^{\rm KK}$ and $T_+(3,1)^{\rm KK'}$. We have also noticed that the 2nd and 3rd pair are brighter, which suggests a larger $P_T$ at these transitions. This is expected because flipping the singlet $S(4,0)$ to the $T_+(3,1)^{\rm K'K'}$ or $T_+(3,1)^{\rm KK}$ states does not require flipping the valley of the transitioning electron (Fig.\ \ref{fig: device}(c) insets) {suggesting the valley states maybe similar between dots.}  From these observations we find valley splittings of 12 $\mu$eV and 33 $\mu$eV but we cannot tell which corresponds to $\Delta_L$ or $\Delta_R$. We take $\Delta_L > \Delta_R$ for convenience in the following discussion. 
	
	Our experiments take place with an electron temperature of around 100 mK, which causes us to load the excited singlet state, $S(3,1)^{\rm K'K'}$, with approximately a 10\% probability.  The energy spectrum of this state is plotted as the red curve in Fig.\ \ref{fig: device}(c). We observe signatures of the degeneracy between this exited singlet state and the 4 triplet states as an additional four sets of curvatures indicated by the red arrows in Fig.\ \ref{fig: device}(d). 
	The curvatures near $B=0$ maps the degeneracy between $S(3,1)^{\rm K'K'}$ and $T_+(3,1)^{\rm K'K'}$. The crossing between $S(3,1)^{\rm K'K'}$ and $T_+(3,1)^{\rm KK}$ and $T_+(3,1)^{\rm KK'}$ leads to parallel curves with offsets of $\Delta_R-\Delta_L = 21 \mu$eV and $\Delta_L = 33 \mu$eV realtive to zero field. The crossing between $S(3,1)^{\rm K'K'}$  and $T_-(3,1)^{\rm K'K}$ gives rise to curvatures with an offset of $\Delta_R =12 \mu$eV relative to zero field and a flipped direction of curvature in response to the magnetic field because it is from the opposite Zeeman branch.

	\begin{figure}
		\includegraphics[width = \columnwidth]{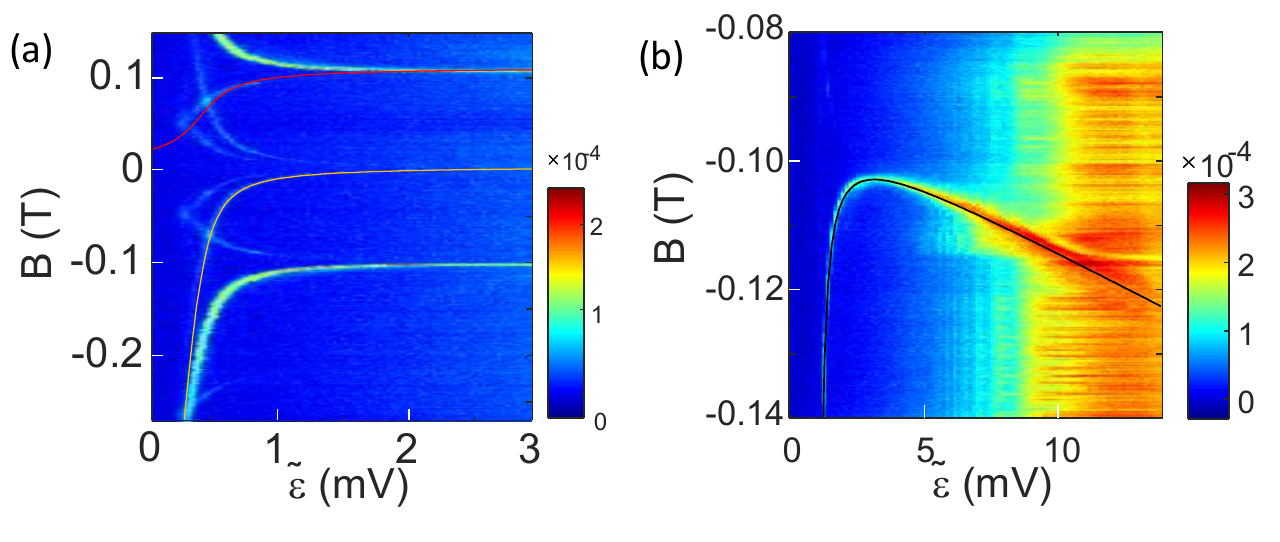}		
		\caption{(a) Fit to the first order spin funnel lines in Fig.\ \ref{fig: device}(d). (b) Funnel measurements when the magnetic field zooms in the second branch and $\tilde{\epsilon}$ extend the over large range to explore the gate dependence of the valley splitting.
		}
		\label{fig: funnel}
	\end{figure}
	
	Figure\ \ref{fig: funnel}(a) overlaps fits for the ground and first excited singlet energies (red and orange curve in Fig.\ \ref{fig: device}(c)) with the funnel data in Fig.\ \ref{fig: device}(d). The fitting finds reasonable agreement between the model and data using the interdot coupling as the only free parameter with a best fit of $t_c = 15$ $\mu$eV.	
	We further utilize this spectroscopy method to measure the gate dependence of $\Delta_{L(R)}$. Fig.\ \ref{fig: funnel}(b) shows the second funnel in greater detail by reducing the range of $E^c_z$ and enlarging the range of $\tilde{\epsilon}$. We fit the curve assuming $\Delta_{R} = \Delta_{R0} + \tilde{\epsilon} \Delta'_{R}$ and find $\Delta'_R=0.3$ meV/V. Similarly we find $\Delta'_L =- 0.09$ meV/V. \cite{SOM} Assuming that $\Delta_{L(R)}$ is only dependent on $P_{1(2)}$ we have $\partial\Delta_{L/R}/\partial V_{P1(2)}$ = 0.3(0.09) meV/V, which is comparable to results reported in previous works \cite{Goswami2007, Friesen2010, Yang2013, Hollmann2020}.

	\section{Tunable two axis control} 
	\label{sec: dBz}
	
	\begin{figure}
		\includegraphics[width=\columnwidth]{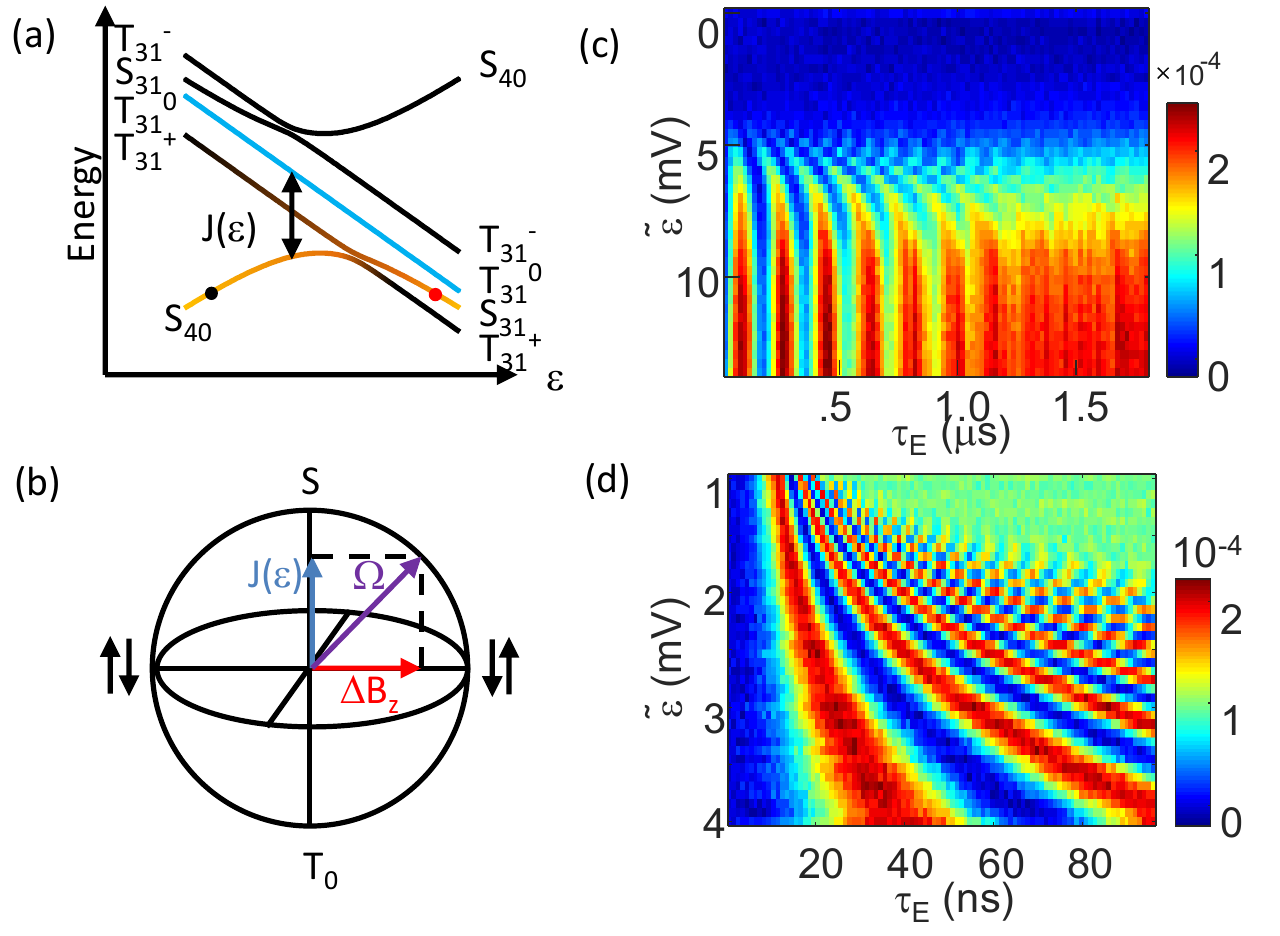}
		\caption{(a) level diagram of relevant states for qubit. The ground singlet (orange) and $T^0$ (blue) states are the base of the singlet-triplet qubit. (b) The Bloch sphere for singlet-triplet qubit. (c-d) the charge readout as a function of $\tilde{\epsilon}$ and exchange time after the $\Delta B_Z$ rotation (c) and the exchange rotation (d).}
		\label{fig: Linear dbz}
	\end{figure}	
		
	Coherent qubit control is explored in the subspace spanned by $S(4,0)$ and $\rm (3,1)^{\rm K'K}$, as demonstrated in Fig.\ \ref{fig: Linear dbz}(a). The energy gaps between the other (3,1) valley states and $\rm (3,1)^{\rm K'K}$ are much larger than the qubit energy scales and can be neglected when considering qubit operations. The ground singlet (orange) and $T_{0}(3,1)^{\rm K'K}$ (blue) constitute the basis of the qubit and the Hamiltonian in this subspace can be approximated by $H = J(\epsilon)\sigma_z + (g_L  B_{zL}-g_R B_{zR})\mu_B\sigma_x$. Here $\sigma_{x,z}$ are the Pauli operators, $g_{L(R)}$ and $B_{zL(R)}$ are the g-factor and external field at left(right) dot. This system can achieve flexible two axis control, as illustrated by the Bloch sphere in Fig.\ \ref{fig: Linear dbz}(b) \cite{Wu2014, Maune2012}. Rotation along the Z axis can be performed near the interdot transition where $J(\epsilon)$ is large and \{$S$, $T_0$\} are the eiganstate of the system Hamiltonian. X axis rotations ($\Delta B_z$ rotations) are achieved at large $\epsilon$ where J($\epsilon$) is negligible and the energy splitting is set by $\mu_B(g_L  B_{zL}-g_R B_{zR})$ and the eigenstates are $|\uparrow\downarrow\rangle$ and $|\downarrow\uparrow\rangle$ \cite{Foletti2009, Wu2014}.
	
	We characterize the $\Delta B_Z$ rotations by initializing the system in $S(4,0)$ and then pulsing the DQD to large $\epsilon$ to turn off $J$, allowing the qubit to rotate around the $\Delta B_Z$ axis into superpositions of $S$ and $T_0$. We then pulse back to $M$ for readout \cite{Foletti2009, Wu2014}. Figure\ \ref{fig: Linear dbz}(c) demonstrates charge readout as a function of exchange time $\tau_E$ and location ($\tilde{\epsilon}$) when we set the field $B_z = 1$ T.  For large $\tilde{\epsilon}>10$ meV, we find an oscillation at a frequency $f_{\rm \Delta Bz} =\mu_B (g_L  B_{zL}-g_R B_{zR})/\hbar =$ 5.5 MHz. The decay in the oscillation amplitude indicates a {$T^*_2 = 1$ $\mu$s} \cite{SOM}. When $\tilde{\epsilon}<10$  mV, the oscillation rate is larger than $f_{\rm \Delta Bz}$ and the amplitude is smaller, indicating a finite $J(\epsilon)$, which contributes to the rotation rate and shifts the angle from the $\Delta B_Z$ axis.  The coherent J rotation can be characterized by a similar process by adding adiabatic ramping before and after the exchange rotation to map S to $|\downarrow\uparrow\rangle$  and $T_0$ to $|\uparrow\downarrow\rangle$ \cite{Petta2004}. The result is presented in Fig.\ \ref{fig: Linear dbz}(d). For $1 < \tilde{\epsilon} < 4$ mV the rotation amplitude is maximized and the rotation rate is strongly dependent on $\tilde{\epsilon}$. Here J is dominant, and the rotation rate is electrically tunable as expected.
	
\begin{figure}
		\includegraphics[width=\columnwidth]{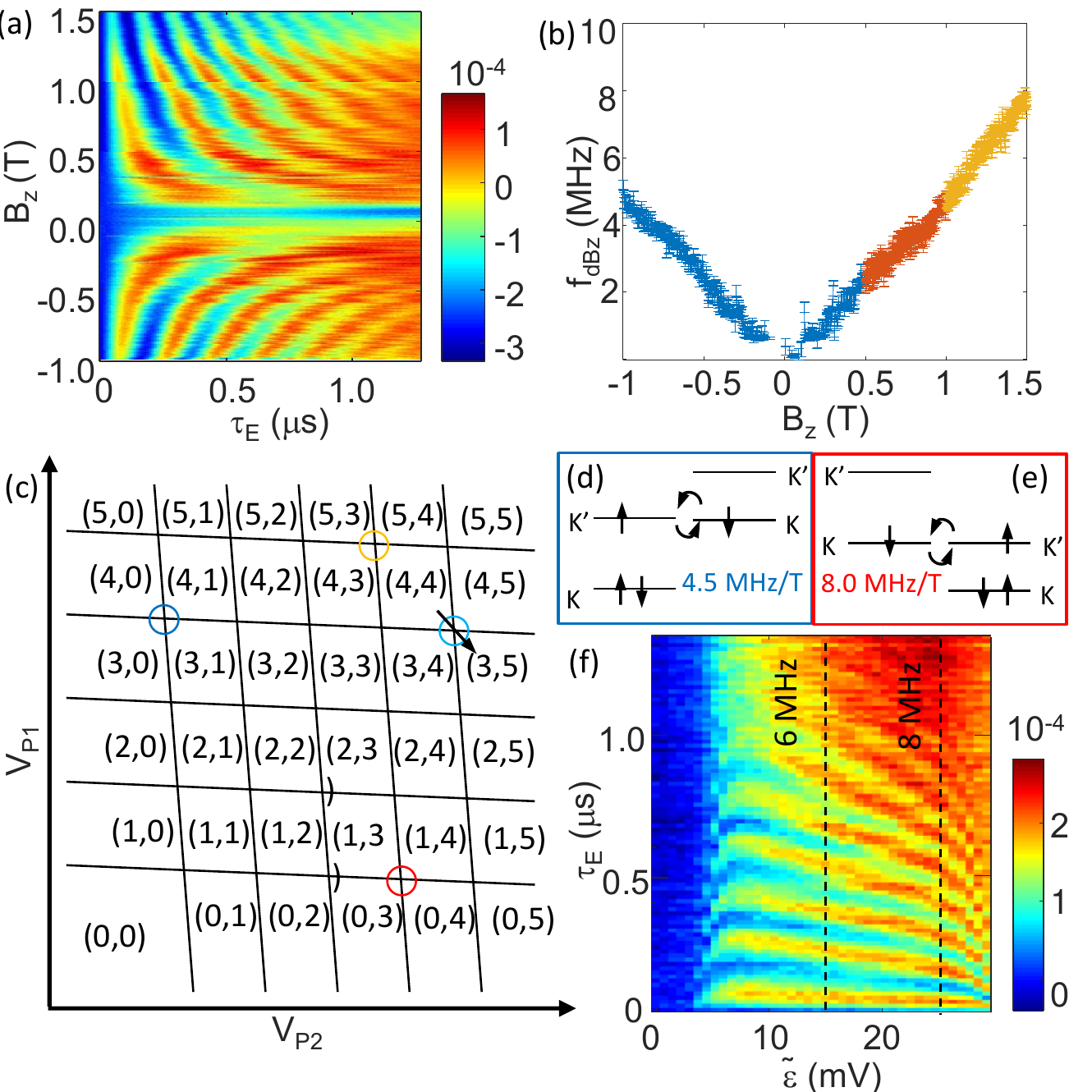}
		\caption{(a) The charge readout as a function of exchange time and magnetic field after a $\Delta B_Z$ rotation with $\tilde{\epsilon} = 10$ mV (b) After $\Delta B_Z$ rotation rate as a function of magnetic field. (c) Schematic of a charge stability diagram. Colored circles labels the transitions that satisfy (4n, 4m)-(4n$\pm$1, 4m$\mp$1). The arrow indicates the direction of $\epsilon$. (d-e) {The ground state transition at (d) (4,0)-(3,1) and (e) (0,4)-(1,3) charge configuration} and the associated $f_{\rm dBz}/B_z$ (f) The charge readout as a function of $\tilde{\epsilon}$ and exchange time after the $\delta B_z$ rotation at (4,4) transition. Dashed lines indicates the range of $\tilde{\epsilon}$ that X axis rotation is dominant and the rotation rate is labeled on top. The gradient of $\Delta B_z$ rotation is 2MHz/10mV.}
		\label{fig: dBz vs gate}
	\end{figure}

	
We emphasize that no micromagnet or other external magnetic field gradient source was added to this device.  To explore the mechanism responsible for the field gradient, we measure $f_{\rm \Delta B_Z}$ as a function of $B_z$.  Figure\ \ref{fig: dBz vs gate}(a) plots charge readout as a function of $\tau_E$ and $B_z$ after a $\Delta B_Z$ rotation at $\tilde{\epsilon} = 10$ mV such that $J(\tilde{\epsilon})$ is negligible. The $\Delta B_Z$ rotation rate is then extracted and plotted as a function of field in Figure\  \ref{fig: dBz vs gate}(b). 
This result is consistent with a difference in the g factor of the two dots of around $\Delta g = 3.8 \times 10^{-4} = 0.02 \% g$ \textcolor{red}. We note that the two dots' ground states occupy different valley states and this $\Delta g$ is consistent with the g factor difference between valleys as previously reported \cite{Kawakami2014, Ferdous2018}. 
	
	This valley introduced $\Delta g$ allows the rate of the $\Delta B_Z$ rotations to be tuned by an external magnetic field. We expect the $\Delta B_Z$ rotation rate would be 14 MHz at 3 T (beyond the current limit of our magnet), comparable to the field gradient generated by a micro-magnet \cite{Wu2014}. This would potentially reduce design complexity for a large array of spin qubits because it eliminates the need for an artificially generated field gradient. 
	
	In order to verify the generality of this phenomena, we measured other (4n, 4m)-(4n$\pm$1, 4m$\mp$1) transitions where the 4 valley spin states are all filled up in one dot and the relevant orbital of the other dot is empty. As previously described, near these transitions the ground states of the (4n, 4m) and (4n$\pm$1, 4m$\mp$1) states occupy different valley states, which introduce a $\Delta g$ to the singlet-triplet qubit Hamiltonian.
	
	Figure\ \ref{fig: dBz vs gate}(c) labels these (4n, 4m)-(4n$\pm$1, 4m$\mp$1) transitions with colored circles in the charge stability diagram. Fig.\ \ref{fig: dBz vs gate}(d) demonstrates the ground states for the (4,0)-(3,1) transition, which is the symmetric configuration compared to the (0,4)-(1,3) transition as demonstrated in Fig.\ \ref{fig: dBz vs gate}(e). At the (1,3)-(0,4) transition, we measure $f_{\rm \Delta Bz}/B_z =$ 8 MHz/T which is almost double the value for the (4,0)-(3,1) transition. $V_{\rm P1}$ and $V_{\rm P2}$ have been changed by 100 mV, which allows us to estimate that the gradient can be tuned at a rate of 35 MHz/(T $\cdot$ V).  The (4,4)-(3,5) transition has the same filling as Fig.\ \ref{fig: dBz vs gate}(d) and (5,3)-(4,4) transition corresponds to Fig.\ \ref{fig: dBz vs gate}(e). Although the two transitions are close in the charge stability diagram, we find a dramatic change in the $\Delta g$. At the (5,3)-(4,4) we find $f_{\rm \Delta Bz}/B_z = 0$ and at the (4,4)-(3,5) transition $f_{\rm dBz}/B_z =$ 5.5 MHz/T at $\tilde{\epsilon} = 15$ mV. 
	
	This drastic change indicates a strong gate dependence of $\Delta g$ . Figure\ \ref{fig: dBz vs gate}(f) demonstrates the $\Delta B_z$ rotation at the (4,4)-(3,5) transition using charge readout as a function of $\tilde{\epsilon}$ and $\tau_E$. Unlike the (4,0)-(3,1) transition where the oscillation rate is a constant once $\tilde{\epsilon} > 10$ mV, we find this frequency increases with $\tilde{\epsilon}$ once $\tilde{\epsilon}>8$ mV at the (4,4)-(3,5) transition.  The oscillation amplitude is maximized in this regime, indicating that $J$ is always negligible such that the rotation is along X axis with the rate only determined by $f_{\rm \Delta Bz}$. At $\tilde{\epsilon} = 15$ mV $f_{\rm \Delta Bz}/B_z = 6$ MHz/T and at $\tilde{\epsilon} = 25$ mV, $f_{\rm \Delta Bz}/B_z$ has increased by 2MHz/T. If we linearly interpolate $f_{\rm \Delta Bz}/B_z$ for $\tilde{\epsilon} <10$ mV, we expect the $f_{\rm \Delta Bz}/B_z=0$ at $\tilde{\epsilon}_0 = -15$ mV. The (5,3)-(4,4) transition is located at $\Delta V_{\rm P1} = -\Delta V_{\rm P2} > 40$ mV from the (4,4)-(3,5) transition and thus $\tilde{\epsilon} <\tilde{\epsilon}_0$. Hence $f_{\rm \Delta Bz}/B_z = 0$ at the (5,3)-(4,4) transition is not completely surprising due to this rapid change in $\Delta g$ at the (4,4)-(3,5) transition. 
		
	The g factor difference between valleys has been explained by spin-orbital coupling in previous works \cite{Kawakami2014, Ferdous2018}. A gate dependent g factor has been predicted for SiGe \cite{Ferdous2018} and observed in Si-MOS system \cite{Veldhorst2015}. The gate dependence of the g factor observed here could enable gate operations that require fast changes in $\Delta B_z$. Compared to the field gradient generated by a micro-magnet, $f_{\rm \Delta Bz}$ by $\Delta g$ can be arbitrarily high with increasing external field. As with the valley splitting, a maximized $\Delta g$ is expected for an atomically flat Si/SiGe interface and high external electrical field. We thus expect a potentially higher $\Delta B_z$ gate fidelity with better substrates and smaller QDs.  
		

	\section{Conclusion}
	
	In this work we have investigated DQD transitions where the electron number is (4n,4m)-(4n$\pm$1,4m$\mp$1). The funnel measurements there provide the valley spectrum in each dot. The extracted valley splitting demonstrates a gate dependence, as observed in previous works.  The ground state in each dot occupies a different valley state which provides a g-factor difference between the two charge states. This g-factor gradient generates a $\Delta B_z$ rotation that is linear to an external magnetic field and is also gate dependent. These two dependencies provide a tunable $\Delta B_z$ rotation that does not require a micro-magent. This would potentially simplify scaling up to large arrays of spin qubits for quantum information processing.
	
	\section{Acknowledgement}
	
	We acknowledge Lisa Edge from HRL Laboratories for the growth and distribution of the Si/SiGe heterostuctures that were used in this experiment. This work was sponsored by the Army Research Office (ARO), and was accomplished under Grant Number W911NF-17-1-0274 and W911NF-17-1-0202. We acknowledge the use of facilities supported by NSF through the UW-Madison MRSEC (DMR-1720415) and the MRI program (DMR-1625348). The views and conclusions contained in this document are those of the authors and should not be interpreted as representing the official policies, either expressed or implied, of the Army Research Office (ARO),or the U.S. Government. The U.S. Government is authorized to reproduce and distribute reprints for Government purposes notwithstanding any copyright notation herein.

\bibliographystyle{apsrev_lyy}
\bibliography{SiGe_ref_v1}
	
\end{document}